\documentclass[conference, letterpaper]{IEEEtran}

\newcommand{\dashboard}[1]{DAS-Dashboard}
\newcommand{\gateway}[1]{GW#1}

\usepackage{amsmath}
\usepackage{mathtools}
\usepackage{dsfont}
\usepackage{paralist}
\usepackage[table]{xcolor}
\usepackage{footnote}
\usepackage{array}
\usepackage{ragged2e}
\usepackage{bm}
\usepackage{textcomp}
\usepackage{hyperref}

\usepackage{subcaption}
\usepackage{mathtools}
\usepackage{algorithmic}

\usepackage{verbatim}

\ifCLASSINFOpdf
  \usepackage[pdftex]{graphicx}
  \graphicspath{{figures/}{plots/}}
\else
\fi

\usepackage{rotating}
\usepackage{multirow}

\begin{document}


%
\title{A Self-Managed Architecture for Sensor Networks Based on Real Time Data 
Analysis}

\author{\IEEEauthorblockN{Gabriel Martins Dias, Toni Adame, Boris Bellalta and Simon 
Oechsner}
\IEEEauthorblockA{~\\Department of Information and Communication Technologies\\ 
Universitat Pompeu Fabra, Barcelona, Spain\\
Email: \{gabriel.martins, toni.adame, boris.bellalta, simon.oechsner\}@upf.edu}
}

\maketitle

\begin{abstract}
Wireless sensor networks (WSNs) have been adopted as merely data producers for years. However, the data collected by WSNs can also be used to manage their operation and avoid unnecessary measurements that do not provide any new knowledge about the environment. The benefits are twofold because wireless sensor nodes may save their limited energy resources and also reduce the wireless medium occupancy. We present a self-managed platform that collects and stores data from sensor nodes, analyzes its contents and uses the built knowledge to adjust the operation of the entire network. The system architecture facilitates the incorporation of traditional WSNs into the Internet of Things by abstracting the lower communication layers and allowing decisions based on the data relevance. Finally, 
we demonstrate the platform optimizing a WSN's operation at runtime, based on different real-time data analysis.

\end{abstract}


\section{Introduction}

Wireless sensor nodes are small computer devices with low production costs, equipped with a radio antenna and sensors capable of sensing one or more environmental parameters~\cite{Akyildiz2002}. As sensor nodes are designed to have low production costs, their computational and energy resources are several orders of magnitude smaller than those of typical workstations. This constraint does not prevent Wireless Sensor Networks (WSNs) to be composed of hundreds of wireless sensor nodes deployed to monitor the environment and engineering structures, track objects, detect meteorological phenomena, among others.

Apart from being used for traditional monitoring tasks, in the last years, WSNs have been incorporated into Internet of Things (IoT) applications, where connected cloud computing services analyze the collected data and consequently trigger reactions~\cite{Bellavista2013}. As sensor nodes usually cannot compute complex algorithms that require a long runtime or significant memory resources, a data analysis performed in sensor nodes may not be as accurate or as fast as those computed in multi-core workstations with high storage and memory capacity.

As a side note, the IoT has a broader scope than traditional WSNs, because it is composed of more powerful devices that can compute complex algorithms, interact with humans and also provide machine-to-machine communication. Indeed, the power of the IoT is not only concentrated in devices and their applications but also, among others, in potential data analytics and interactions between different device types. Therefore, while early (traditional) WSNs performed simple data collection tasks and were merely considered as data providers, sensor nodes can be nowadays benefited by intelligent data applications, such as the optimization of the sensors' sampling interval based on data predictions~\cite{Dias2016}. By way of illustration, in the Entomatic project\footnote{http://entomatic.upf.edu}, wireless sensor nodes periodically report information on pest population density and environmental parameters, such as temperature and relative humidity. The data analysis results can be used to spray pesticides intelligently, i.e., only when--and where--necessary.

Our contribution is an architecture that relies on a robust entity to exploit the asymmetry in WSN devices' capacities when computing data algorithms. First, it overcomes a common restriction imposed in other platforms for WSNs~\cite{turon2005mote,ruzzelli2008octopus}, where WSN owners have to upload new applications manually whenever they decide to take decisions based on sophisticated data processing techniques computed outside the WSNs. At the same time, our architecture completes other IoT platforms that integrate a user interface to visualize and analyze the collected data~\cite{nimbits2016,realtimeio2016}, because it offers means for WSNs improving their operation according to the knowledge acquired in data analysis. In this paper, we consider a set of use cases to illustrate these features.

\section{Proposed Architecture}
\label{section:architecture}

We propose an architecture that exploits the sensor nodes hardware limitations and the physical distance from data origin to the central server. Our platform fits the main principles of data science related with sensed data: its collection; description; storage; maintenance; discovery; visualization; and analysis. As the platform stores and publishes collected sensed data, it is possible to visualize measurements and other collected values, as well as reproduce IoT scenarios to optimize data acquisition and its further analysis.

The data analysis can rely on complex computations to extract knowledge from the collected data and provide reliable services at a low cost~\cite{7123563}. In this architecture, they can be delivered as a service to WSNs, such that sensor nodes will be able to apply the knowledge in their favor, e.g., changing their operation to report measurements more often and detail the variations in the environment.

\subsection{Components}

\begin{figure}[t]
        \centering
	\includegraphics[width=0.45\textwidth]{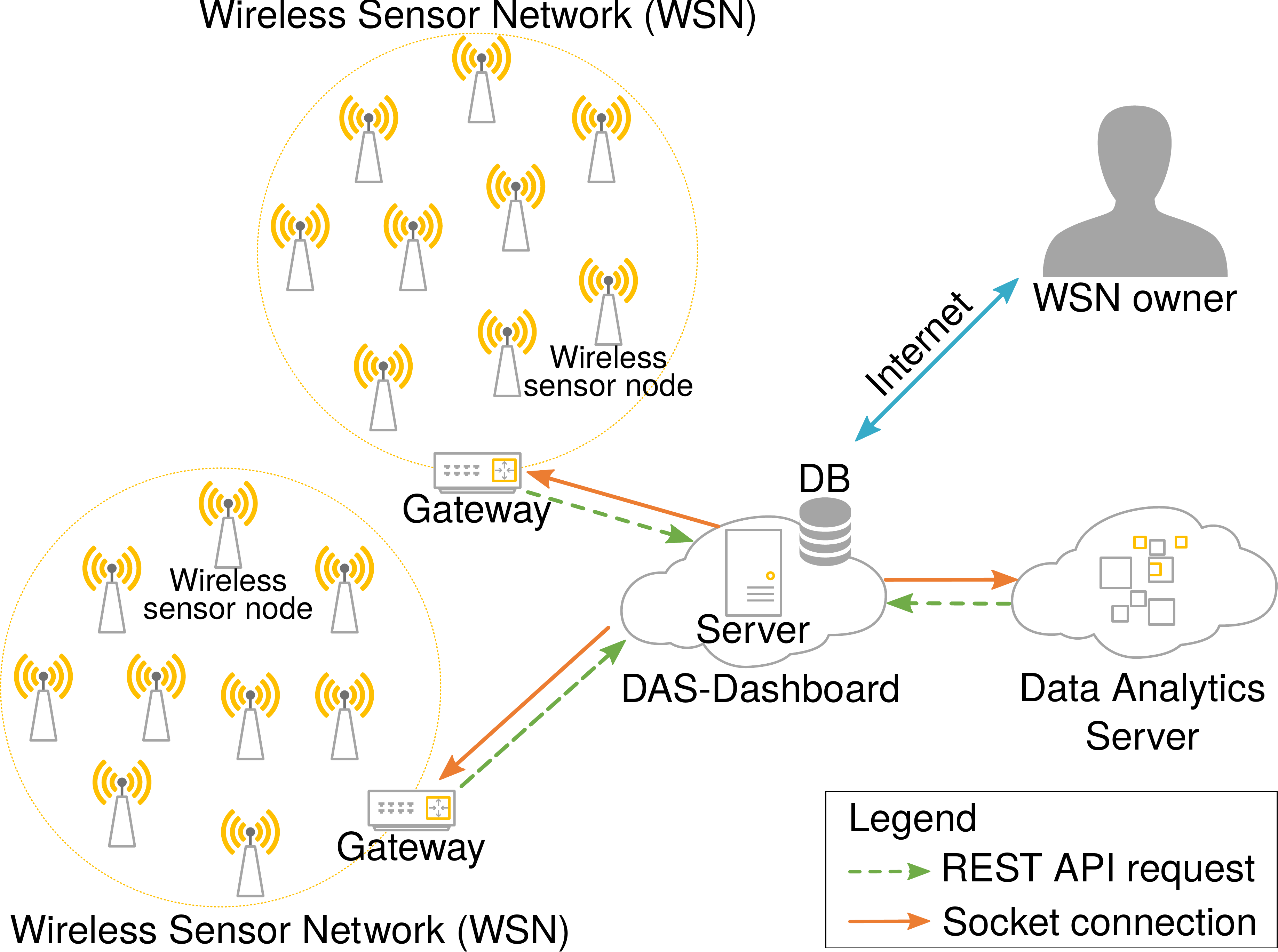}
        \caption{System Architecture}
        \label{fig:system-architecture}
\end{figure}

We consider a typical IoT environment composed by several WSNs with ordinary wireless sensor nodes reporting to predetermined sinks, called as \emph{Gateways} (\gateway{s}) in Figure~\ref{fig:system-architecture}. The proposed architecture is centered in the \emph{Data Analytics for Sensors 
Dashboard} (\dashboard{}) and consists of interconnected components that can exchange information with trusted entities, eventually from different domains. The components of this architecture are:

\subsubsection{Wireless sensor nodes}
Taking advantage of their proximity to the data origin, they perform default sensing tasks in their deployment area and transmit their measurements via radio to a \gateway{}.

\subsubsection{Gateways}
They forward the gathered information to the central server and disseminate occasional instructions and updates to wireless sensor nodes. \gateway{s} are the link between ordinary wireless sensor nodes and the \dashboard{}, using a point-to-point connection over the Internet or a local network.

\subsubsection{\dashboard{}}
The central component of this architecture has three primary responsibilities: collecting, storing and publishing data transmitted by wireless sensor nodes. The ability to collect data requires a direct communication with the WSN and is fundamental for the other two responsibilities. Storing the collected data in a database (DB) allows further access to historical information, besides providing data visualization to network owners and other users. Finally, communicating data to other systems allows the dashboard to outsource data processing, such as filtering and analyzing the collected data and, especially, predicting future measurements.

\subsubsection{Data Analytics Server}
The Data Analytics Server can process computationally expensive real-time analysis over data. To do that, it can rely on external data resources, such as public services and other databases via the Internet. Indeed, tasks processed by the Data Analysis Server could not be assigned to sensor nodes, due to their constrained hardware and limited communication with external networks.

\subsection{Communication with external components}

The communication between \dashboard{} and any external component follows a standard: external data (coming from \gateway{s} and the Data Analytics Server) is received via APIs (GET or POST HTTP requests); and outbound data is announced in the form of events via socket connections to previously registered services that keep listening for updates. If connected to the Internet, the \dashboard{} may make data publicly available to remote access via an online web-based platform. Hence, besides storing data and managing users access, the system architecture supports data analytics algorithms and allows the \dashboard{} to change sensors' operations according to the data they have previously reported. For example, sensor nodes reconfiguration may be done via peer-to-peer or multicast communication, based on the analysis published by the \dashboard{}.

\section{Self-Management Demo}

In this demonstration, we show that our architecture can exploit the link between the \dashboard{} and a WSN, by analyzing the measured data and adjusting the operation of sensor nodes according to the analysis outcomes. Our WSN is composed by one \gateway{} and four sensor nodes that work independently of each other, representing use cases in which sensor nodes monitor parameters from indoor and outdoor environments.

The \dashboard{} is deployed in a well-dimensioned machine without energy nor performance constraints, with reliable Internet connection and direct access to the other architecture components. A Web User Interface provides the user a means of visualizing the collected data, the sensor nodes' location in a map, occasional failures that may be reported and results from the data analysis, performed by an external server. The Data Analytics Server is developed in R programming language and may run two types of analysis: \emph{data relevance assessment} and \emph{forecasting}. In the following, we describe each analysis procedure, illustrating the interconnection between a WSN, the \dashboard{} and the Data Analytics Server in a real deployment.

\subsection{Data relevance assessment}

In this analysis, data is used to calculate the importance of a sensor in the measurements reported by the whole WSN. To achieve that, it may be possible to use external information, such as the time of the day and the temperature reported by an online weather service. As a result, if the data analysis is about the relevance of the data generated by sensor nodes, the output is a suggestion about updating the sensor nodes' operation to sample more (or less) often.

\subsubsection{Single rule demo}

\begin{figure}[t]
	\centering
	\includegraphics[width=0.38\textwidth]{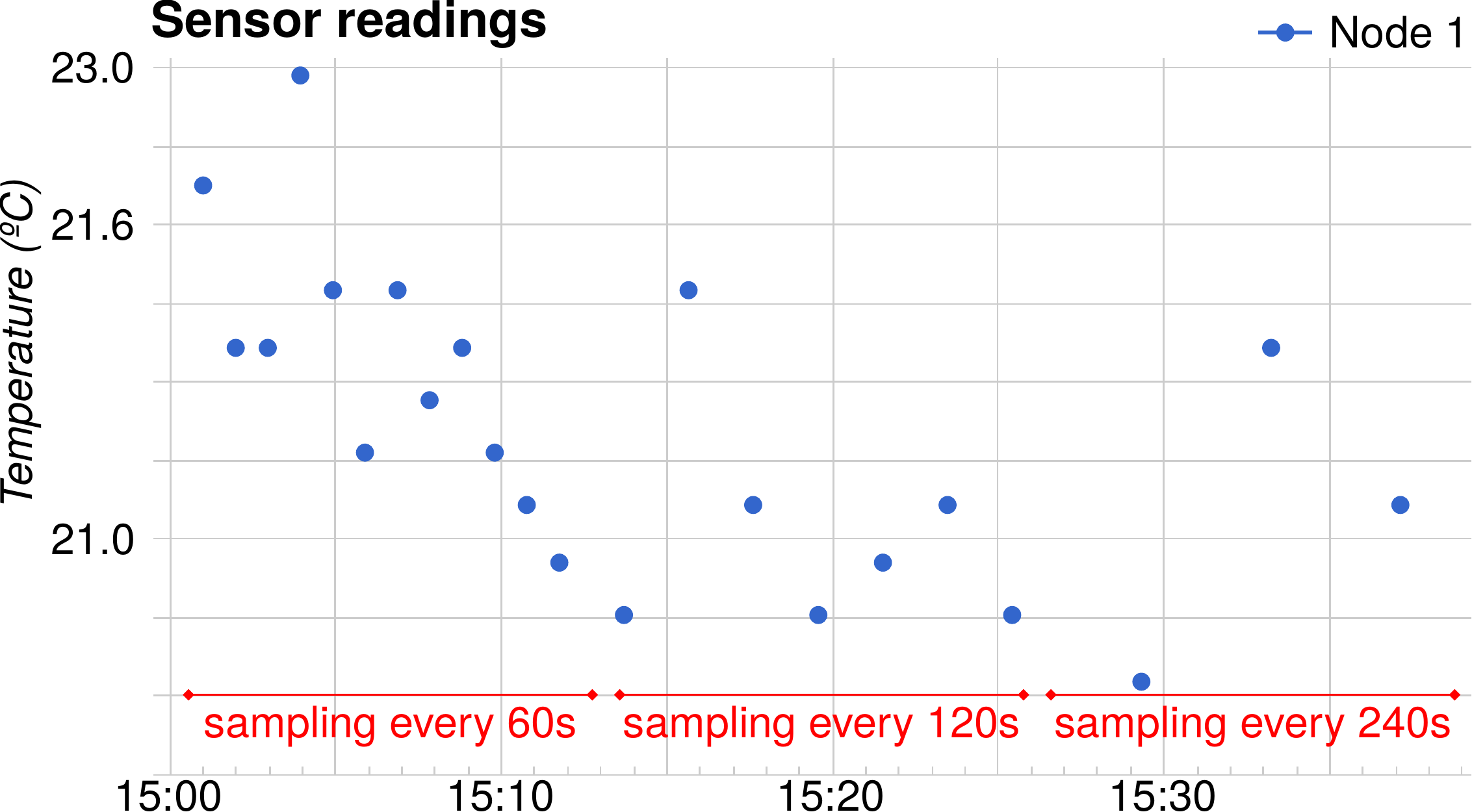}
	\caption{In this example, the sampling interval is updated every $12$ minutes, and it varies between $60$, $120$ and $240$ seconds.}
	\label{fig:adaptive-sampling}
\end{figure}

One example of this type of data analysis is a single rule defining the sampling interval according to the time of the day. For instance, if it is a working hour, sensor nodes placed in an office have to measure the temperature once every $5$ minutes; otherwise, they can be set to measure it less often (e.g., every $30$ minutes). Hence, the WSN would report more detailed information when people are in its surroundings because their presence may impact the temperature and also because other systems (such as the air conditioning and personal computers) work only during working hours, provoking higher variations and sharper changes. Figure~\ref{fig:adaptive-sampling} depicts an image from the \dashboard{}, illustrating a rule that changes the sampling interval every $12$ minutes.

\subsubsection{External information demo}

Another example is the use of the Internet access to compare the data reported by sensor nodes with the temperature values reported by a weather service. If the values coincide sufficiently, the \dashboard{} stores in the database the forecast values from the weather service and reduces the sensor nodes' sampling interval to save their battery. Otherwise, sensor nodes keep measuring as often as possible to inform the temperature values accurately to the WSN owner. The support of a reliable external forecasting service can be justified by the high complexity of forecasts provided by its supercomputers, which are unfeasible to be locally performed without additional--and significant--costs.

\subsection{Forecasting}

\begin{figure}[t]
	\centering
	\includegraphics[width=0.45\textwidth]{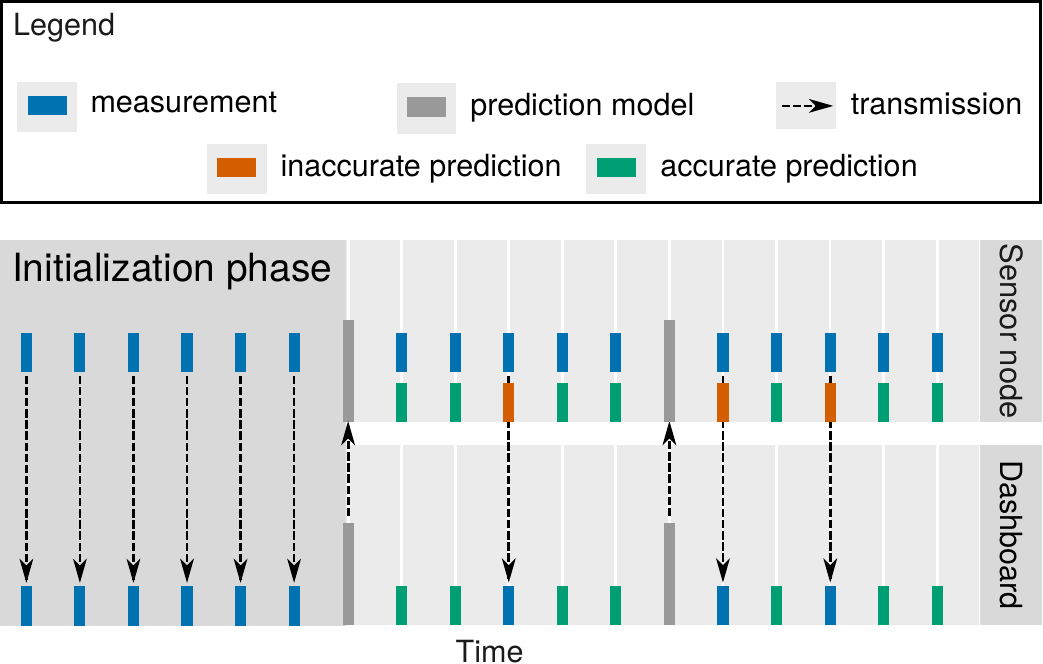}
	\caption{In a DPS, a measurement is transmitted only if its
	forecast is inaccurate. The \dashboard{} may be responsible for 
transmitting new prediction models every time interval after the 
initialization phase.}
	\label{fig:timeline-dps-ch}
\end{figure}

The strategy known as Dual Prediction Scheme (DPS) relies on forecasts of values that will be measured by the wireless sensor nodes~\cite{Dias2016b}. After predicting the future measurements, the \dashboard{} informs the sensor nodes which specific values are expected to be measured, as illustrated in Figure~\ref{fig:timeline-dps-ch}. If the actual measurements match to the forecasts (or do not differ by more than an accepted threshold), they are not transmitted, saving sensor nodes' batteries and reducing the wireless medium occupancy.

\subsubsection{Autoregressive Integrate Moving Average demo}

The AutoRegressive Integrate Moving Average (ARIMA) is a technique used to forecast values which can accurately predict up to $20$ future measurements and reduce up to $50\%$ of the number of transmissions~\cite{Dias2016}. Due to its high computational complexity, an ARIMA model computation might waste sensor nodes' battery without providing enough savings to compensate it. Therefore, even though the ARIMA method is not suitable to the simplest wireless sensor nodes, in this architecture we can exploit the computational asymmetry of WSN devices and avoid unnecessary data transmissions.

\section{Conclusion}
In this paper, we present a system architecture that enables the integration of device types with different computing powers and capacities. To demonstrate its application in a real-world use case, we apply the obtained results in real-time data analysis to optimize a WSN's operation and visualize its evolution in a Web User Interface.

\section*{Acknowledgment}

This work has been partially supported by the Spanish Government through the 
project TEC2012-32354 (Plan Nacional I+D), by the Catalan Government 
through the project SGR-2014-1173 and by the European Union through the 
project FP7-SME-2013-605073-ENTOMATIC.

\bibliographystyle{IEEEtran}
\bibliography{IEEEabrv,bibliography}

\end{document}